\documentclass[aps,prd,twocolumn,showpacs,floatfix,superscriptaddress,
preprintnumbers,nofootinbib]{revtex4}
\usepackage{graphicx,amssymb}

\begin{document}

\title{Possible Types of the Evolution of Vacuum Shells
around the de Sitter Space}
\author{V. I. Dokuchaev}
 \email{dokuchaev@inr.ac.ru}
 \affiliation{Institute for Nuclear Research of the Russian
 Academy of Sciences, 60-letiya Oktyabrya pr. 7a, Moscow, 117312
 Russia}
\author{S. V. Chernov}
 \email{chernov@td.lpi.ru}
 \affiliation{Institute for Nuclear Research of the Russian
 Academy of Sciences, 60-letiya Oktyabrya pr. 7a, Moscow, 117312
 Russia}
 \affiliation{Lebedev Physical Institute of the Russian Academy of
 Sciences, Leninskii pr. 53, Moscow, 119991 Russia}

\begin{abstract}
All possible evolution scenarios of a thin vacuum shell surrounding the spherically symmetric
de Sitter space have been determined and the corresponding global geometries have been
constructed. Such configurations can appear at the final stage of the cosmological phase
transition, when isolated regions (islands) of the old vacuum remain. The islands of the old
vacuum are absorbed by the new vacuum, expand unlimitedly, or form black holes and wormholes
depending on the sizes of the islands as well as on the density and velocity of the shells
surrounding the islands.
\end{abstract}


\maketitle

The dynamical evolution of a self-gravitating old vacuum island (bubble) appearing between
growing and intersecting new-vacuum islands in the process of the cosmological phase transition
in the early universe is analyzed. The inner and outer parts of the bubble are simulated by the
de Sitter and Schwarzschild metrics, respectively. It is also assumed that the domain wall
separating the outer and inner parts of the bubble is a thin shell. Various particular cases of
this problem were considered in many works on cosmological phase transitions (see, e.\,g..,
early works \cite{Kirznits,KobzOkunVoloshin,ColemanLuccia,Sato}), which can give rise to the
formation of black holes and various types of wormholes
\cite{Sato,BerKuzTkach,IpserSikivie,Aurilia,BlGuenGuth,FrolMarMuk}. In this work, all possible
evolution scenarios of a thin vacuum shell surrounding the spherically symmetric de Sitter
space are determined and the corresponding global geometries are constructed. The determination
of possible evolution scenarios for vacuum bubbles can serve as a basis for calculations of the
probabilities of forming wormholes and primary black holes in detailed models of phase
transitions.

In the thin-shell formalism \cite{Israel}, the equations of motion of the vacuum bubble can be
written in the form \cite{BerKuzTkach,BlGuenGuth}
\begin{equation}
 4\pi S \!=\!
 \frac{\sigma_{\rm in}}{\rho}\sqrt{\dot{\rho}^2\!+\!1\!-\!
 \frac{8\pi}{3}\rho^2\varepsilon}-
 \frac{\sigma_{\rm out}}{\rho}\sqrt{\dot{\rho}^2\!+\!1\!-\!
 \frac{2m}{\rho}},
 \label{evol}
\end{equation}
where $\rho(\tau)$ is the bubble radius, $\tau$ is the observer time on the shell, $S$ is the
surface energy density of the bubble on the shell, $\varepsilon$  is the energy density inside
the shell, $m$ is the total Schwarzschild mass of the shell, and $\sigma_{\rm in,out}=\pm1$.
For subsequent analysis, it is convenient to represent this equation of motion in the form of
the equation for the effective energy, $\dot{\rho}^2+U(\rho)=0$, where the effective potential
has the form
\begin{equation}
 U(\rho) = 1-\!
 \left[\frac{\varepsilon(1+e)}{3S}\right]^2\!\rho^2\!-
 \!\left(\!1-\!\frac{1}{e}\right)\frac{m}{\rho}-
 \frac{m^2}{16\pi^2S^2\rho^4},
\end{equation}
In this case,
\begin{eqnarray}
 \sigma_{in} &=&
 sign\left[m-\frac{4\pi}{3}\rho^3\varepsilon(1-e)\right]; \\
 \sigma_{out} &=&
 sign\left[m-\frac{4\pi}{3}\rho^3\varepsilon(1+e)\right].
\end{eqnarray}
where $e=6\pi S^2/\varepsilon$.

This problem involves four characteristic radii: two gravitational radii
\begin{equation}
 \rho_{\rm bh,1}=\sqrt{\frac{3}{8\pi\varepsilon}}
 \quad \mbox{è} \quad \rho_{\rm bh,2}=2m,
\end{equation}
as well as two radii
\begin{equation}
 \rho_{1}\!=\!
 \left[\frac{3}{4\pi}\frac{m}{\varepsilon(1-e)}\right]^{1/3}
  \quad \mbox{è} \quad
 \rho_{2}\!=\!\left[\frac{3}{4\pi}\frac{m}{\varepsilon(1+e)}\right]^{1/3},
\end{equation}
at which $\sigma$ changes sign. Note that the radius v exists only for $e<1$, i.\,e., for
$\varepsilon>6\pi S^2$. The behavior of the solutions of evolution equation (\ref{evol})
depends on the relative values of the four characteristic radii $\rho_{bh,1}$, $\rho_{bh,2}$,
$\rho_{1}$ and  $\rho_{1}$, and on the presence or absence of the intersection of the potential
with the $U=0$ axis. Let us consider successively all the possible combinations by analogy with
the analysis performed in \cite{BlGuenGuth}.

The condition of the intersection of the potential $U(\rho)=0$ with the $U=0$ axis is
determined by the mass parameter
\begin{equation}
 {m}_{0} =
 \frac{y^{2}}
 {\left\{\!\left[2\pi S\!\left(1\!+\!\frac{1}{e}\right)\right]^2y^2\!+\!
 \left(1\!-\!\frac{1}{e}\right)y\!+\!(4\pi S)^{-2}\right\}^{3/2}},
\end{equation}
where
\begin{equation}
 y =
 \frac{1-\frac{1}{e}+\sqrt{(1-\frac{1}{e})^2+
 8(1+\frac{1}{e})^2}}
 {\left[4\pi S\left(1+\frac{1}{e}\right)\right]^2}.
\end{equation}
The potential intersects the $U=0$ axis only if $m<m_{0}$. It can be easily shown that the
second derivative of the potential is always negative. This means that the stable equilibrium
position is absent in this problem. The case $m<m_{0}$ is divided into subcases. If $m<m_{1}$,
where
\begin{equation}
 m_{1}
 =\left(\frac{3}{32\pi\,\varepsilon}\right)^{1/2}(1-e)<m_{0},
\end{equation}
the four radii satisfy the inequalities $\rho_{\rm bh,2}<\rho_{2}<\rho_{1}<\rho_{\rm bh,1}$ (if
$\rho_{1}$ exists). The radii $\rho_{1}$ and $\rho_{2}$ are in the forbidden region and the
shell does not intersect them in this case. If $e<1$ (i.\,e., $\varepsilon>6\pi S^2$) and the
shell moves from infinity with a negative velocity, i.e., contracts, then the shell reaches the
stop point, is reflected, and begins to expand infinitely. This solution corresponds to case
$C$ from \cite{BlGuenGuth}. This solution corresponds to Carter–Penrose diagram (a) and
embedding diagram (i) in the Fig.~1. According to these diagrams, the shell moves in the
space-time region $R_{-}$ of the outer metric and forms a wormhole \cite{BerKuzTkach}.

Carter–Penrose diagram (b) and embedding diagram (j) in the Fig.~1 correspond to the case $e>1$
and motion of the shell from infinity with a negative velocity. In this case, the shell also
moves in the region $R_{-}$ of the outer metric and forms a wormhole \cite{BerKuzTkach}.

If the shell moves with a positive velocity (expands) from the origin, it reaches the stop
point, is reflected, and begins to contract. In this case, the shell evolution corresponds to
solution ``$A$'' from \cite{BlGuenGuth} and to diagrams (c) and (k) in the Fig.~1. In this
case, the shell does not intersect the region $R_{-}$ and, therefore, a black hole, rather than
a wormhole, is finally formed \cite{BerKuzTkach}.

The next case is $m_{1}<m<m_{2}$, where
\begin{equation}
 m_{1}<m_{2}=
 \left[\frac{32\pi}{3}\varepsilon(1+e)\right]^{-1/2}<m_{0}.
\end{equation}
In this case, the four radii satisfy the inequalities $\rho_{\rm bh,2}<\rho_{2}<\rho_{\rm
bh,1}<\rho_{1}$. The radius $\rho_{2}$ is in the forbidden region and the shell does not
intersect it. In this case, if the shell moves from infinity with a negative velocity, it
reaches the stop point, is reflected, and begins to expand. This solution is solution ``$D$''
from \cite{BlGuenGuth}. This solution corresponds to Carter–Penrose diagram (b) and embedding
diagram (j) in the figure. In this case, the shell intersects the region $R_{-}$ and forms a
wormhole \cite{BerKuzTkach}. If the shell initially moves with a positive velocity from the
origin, this is case ``$A$'' from \cite{BlGuenGuth} and corresponds to diagrams (c) and (k) in
the Fig.~1. In this case, the black hole is finally formed.

\begin{figure}%[t]
\begin{center}
\includegraphics[angle=0,width=0.48\textwidth]{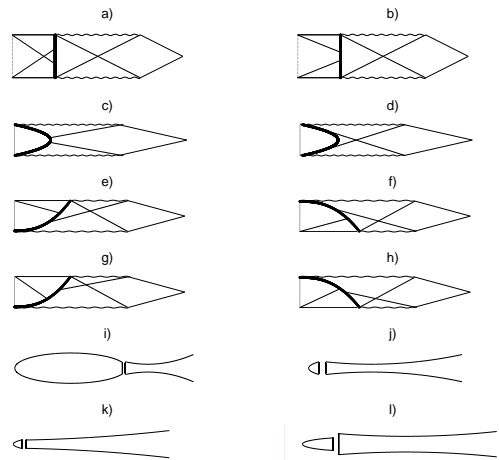}
\end{center}
\caption{Carter–Penrose diagrams and embedding diagrams describing all possible types of the
global symmetry around the de Sitter space.} \label{fig1}
\end{figure}

Correspondingly, when $m_{2}<m<m_{0}$, the potential intersects the $U=0$ axis. In this case,
the four characteristic radii satisfy the inequalities $\rho_{2}<\rho_{\rm bh,2}<\rho_{\rm
bh,1}<\rho_{1}$. The case where the shell moves from infinity with a negative velocity reduces
to the case considered above. This is case ``$D$'' from \cite{BlGuenGuth}. If the shell moves
with a positive velocity from the origin, this is case ``$B$'' from \cite{BlGuenGuth}. This
case corresponds to diagrams (d) and (j). In this case, the shell intersects the region $R_{-}$
of the outer metric and forms a wormhole \cite{BerKuzTkach}.

The final possible case is $m>m_{0}$. In this case, the potential does not intersect the $U=0$
axis and there is no reflection point in the solution. This is case ``$E$'' from
\cite{BlGuenGuth}. We consider this case in more detail, because it was not analyzed in
\cite{BlGuenGuth}. In this case, two subcases are possible. If $m_{0}<m<m_{3}$, where
$m_{3}=(32\pi\varepsilon/3)^{-1/2}$, the four characteristic radii satisfy the inequalities
$\rho_{2}<\rho_{\rm bh,2}<\rho_{\rm bh,1}<\rho_{1}$. If the shell is initially near the origin
and its velocity is positive, it expands infinitely as follows from diagrams (e) and (j). In
this case, the shell intersects the region $R_{-}$ of the outer metric and, therefore, forms a
wormhole \cite{BerKuzTkach}. If the shell is initially at infinity and its velocity is
negative, nothing prevents it from contracting. This case corresponds to diagrams (f) and (j).
It is seen that the shell finally intersects the region $R_{-}$ of the outer metric and forms a
wormhole \cite{BerKuzTkach}.

The next possible case depends on the relation between $\varepsilon$ and $6\pi S^2$.
If$\varepsilon<3\pi S^2(1+\sqrt{5})$, then $m_{4}< m_{5}$, where
\begin{equation}
 m_{4}\!=\!\!
 \left(\!\frac{32\pi\varepsilon}{3}\!\right)^{\!-1/2}\!\!\!(1+e), \;
 m_{5}\!=\!\!
 \left[\frac{32\pi\varepsilon}{3}(1-e)\right]^{-1/2}\!\!.
\end{equation}
When $m_{3}<m<m_{4}$, the four characteristic radii satisfy the inequalities
$\rho_{2}<\rho_{\rm bh,1}<\rho_{\rm bh,2}<\rho_{1}$. This case corresponds to Carter–Penrose
diagram (g) for an expanding shell and diagram (h) for a contracting shell. Corresponding
embedding diagram (l) is shown only conditionally, because the existence regions of the de
Sitter and Schwarzschild embedding diagrams in this case do not overlap. Correspondingly, when
$m_{4}<m<m_{5}$, the four characteristic radii satisfy the inequalities $\rho_{\rm
bh,1}<\rho_{2}<\rho_{\rm bh,2}<\rho_{1}$. In this case, the Carter-Penrose diagram remains
unchanged. In the final case $m>m_{5}$, the four characteristic radii satisfy the inequalities
$\rho_{\rm bh,1}<\rho_{2}<\rho_{1}<\rho_{\rm bh,2}$ and the corresponding diagrams also remain
unchanged. If $\varepsilon>3\pi S^2(1+\sqrt{5})$, then $m_{4}>m_{5}$ and, therefore, $\rho_{\rm
bh,2}$ and $\rho_{1}$, rather than $\rho_{2}$ and $\rho_{\rm bh,1}$, first change their places.
The qualitative pattern of the dynamical evolution does not change in this case.

In conclusion, we note that the total shell mass m that makes it possible to determine and
systematize all possible global geometries of the evolving shell around the de Sitter space is
determined by the energy density $\varepsilon$ inside the shell and the surface energy density
$S$ on the shell, as well as depends implicitly on the initial radius and velocity of the shell
at the instant of bubble formation.

\acknowledgments

This work was supported by the Russian Foundation for Basic Research (project nos. 06-02-16029
and 06-02-16342).


\begin{thebibliography}{99}

\bibitem{Kirznits} D. A. Kirzhnits, Pis'ma Zh. Eksp. Teor. Fiz. 15, 745
(1972) [JETP Lett. 15, 529 (1972)]; D. A. Kirzhnits and A. D. Linde, Phys. Lett. B 42B, 471
(1972).

\bibitem{KobzOkunVoloshin} I. Yu. Kobzarev, L. B. Okun, and M. B. Voloshin,
Yad.Fiz.. {\bf 20}, 1229 (1974) [Sov. J. Nucl. Phys. 20, 644
(1975)].

\bibitem{ColemanLuccia} S. Coleman, F. D. Luccia, Phys.~Rev.~{\bf
D15}, 2929 (1977); {\sl ibid.} {\bf D21}, 3305 (1980).

\bibitem{Sato} K. Sato, M. Sasaki, H. Kodama, and K. Maeda,
Prog. Theor. Phys. {\bf 65}, 1443 (1981).

\bibitem{Israel} W. Israel, Nuovo Cimento. {\bf44B}, 1 (1966);
{\sl ibid.} {\bf48B}, 463 (1967).

\bibitem{BerKuzTkach} V. A. Berezin, V. A. Kuzmin, and I. I. Tkachev, Zh. Eksp.
Teor. Fiz. {\bf86}, 785 (1984); {\sl ibid.} Pis'ma Zh. Eksp. Teor. Fiz. {\bf41}, 446 (1985);
{\sl ibid.} Phys.~Rev.~{\bf D36}, 2919 (1987); {\sl ibid.} Phys.~Lett.~{\bf B120}, 91 (1983).

\bibitem{IpserSikivie} J. Ipser, P. Sikivie, Phys.~Rev.~{\bf D30},
712 (1984).

\bibitem{Aurilia} A. Aurilia, G. Denardo, F. Legovivni, E. Spallucci,
Phys.~Lett. {\bf B147}, 258 (1984); Nucl. Phys. {\bf B252}, 523
(1985).

\bibitem{BlGuenGuth} S.\,K. Blau, E.\,I. Guendelman, A.\,H. Guth,
Phys. Rev. {\bf D35}, 1747 (1987).

\bibitem{FrolMarMuk} V.\,P. Frolov, M.\,A. Markov, V.\, F. Mukhanov,
Phys. Rev. {\bf D41}, 383 (1990).

\end{thebibliography}
\end{document}